\documentclass[aps]{revtex4}

\usepackage[dvips]{graphicx}
\usepackage{bm}

\begin{document}

\title{Noise-induced synchronization and clustering in ensembles\\ of
  uncoupled limit-cycle oscillators}

\author{Hiroya Nakao, Kensuke Arai, and Yoji Kawamura}

\affiliation{Department of Physics, Kyoto University, Kyoto 606-8502,
  Japan}

\date{\today}

\begin{abstract}
  We study synchronization properties of general uncoupled limit-cycle
  oscillators driven by common and independent Gaussian white noises.
  Using phase reduction and averaging methods, we analytically derive
  the stationary distribution of the phase difference between
  oscillators for weak noise intensity.
  We demonstrate that in addition to synchronization, clustering, or
  more generally coherence, always result from arbitrary initial
  conditions, irrespective of the details of the oscillators.
\end{abstract}

\maketitle

Noise-induced synchronization is widely observed in various
experimental systems ranging from neurons to lasers~\cite{NoiseSync}.
From the theoretical standpoint, after several pioneering
studies~\cite{Num}, significant progress has been made by utilizing
the phase reduction method for limit
cycles~\cite{Teramae,Goldobin,Nakao}.
These works generally proved that when the limit-cycle oscillators are
driven by a sufficiently weak common additive noise, the Lyapunov
exponent of the synchronized state averaged over the noise
distribution always becomes negative, namely, the synchronized state
is statistically stabilized.
However, these works are still incomplete as the Lyapunov exponent
only characterizes local stability and do not describe global
behavior of the oscillators.
Also, effects of multiplicative common noises and non-vanishing
additional independent noises remain unclarified.
In this letter, we analyze this phenomenon in more detail from an
alternative perspective by adopting phase reduction and averaging
methods to many-body stochastic dynamical equations describing a
general class of limit-cycle oscillators driven by common and
independent noises, which yields global characterization of their
synchronization properties.

We consider the following Langevin equations describing an ensemble of
$N$ uncoupled identical oscillators driven by common and independent
noises:
\begin{equation}
  \dot{{\bm X}}^{(\alpha)}(t) =
  {\bm F}({\bm X}^{(\alpha)})
  + \sqrt{D} {\bm G}({\bm X}^{(\alpha)}) {\bm \xi}(t)
  + \sqrt{\epsilon} {\bm H}({\bm X}^{(\alpha)}) {\bm \eta}^{(\alpha)}(t),
  \label{lan0}
\end{equation}
for $\alpha = 1, \cdots, N$, where ${\bm X}^{(\alpha)}(t) \in
{\boldsymbol R}^{M}$ represents the state of the $\alpha$-th
oscillator at time $t$, ${\bm F}({\bm X}^{(\alpha)}) \in {\boldsymbol
  R}^{M}$ its individual dynamics, ${\bm \xi}(t) \in {\boldsymbol
  R}^{M}$ the external noise common to all oscillators, and ${\bm
  \eta}^{(\alpha)}(t) \in {\boldsymbol R}^{M}$ the external noise
added independently to each oscillator.
${\bm \xi}(t)$ and ${\bm \eta}^{(\alpha)}(t)$ are assumed to be
independent, identically distributed zero-mean Gaussian white noises
of unit intensity and correlation functions given by $\langle
\xi_{i}(t) \xi_{j}(s) \rangle = \delta_{i,j} \delta(t-s)$, $\langle
\eta^{(\alpha)}_{i}(t) \eta^{(\beta)}_{j}(s) \rangle = \delta_{\alpha,
  \beta} \delta_{i,j} \delta(t-s)$, and $\langle \xi_{i}(t)
\eta^{(\alpha)}_{j}(s) \rangle = 0$ (the subscript $i$ or $j$ denotes
the vector component).
The parameters $D$ and $\epsilon$ control their intensities.
The ${\bm R}^{M \times M}$ matrices ${\bm G}({\bm X}^{(\alpha)})$ and
${\bm H}({\bm X}^{(\alpha)})$ represent the coupling of the oscillator
to both types of the noises, which are assumed to be smooth functions
of ${\bm X}^{(\alpha)}$.
We interpret these Langevin equations in the Stratonovich sense,
namely, we consider the white noise as the limit of colored noise with
vanishingly small correlation time.

We assume that (i) each oscillator obeys the same dynamics, with a
single stable limit cycle ${\bm X}_0(t)$ in its phase
space~\cite{footnote:dispersion}, and that (ii) noises of both types
are sufficiently weak, so that phase
reduction~\cite{Kuramoto,Ermentrout,Izhikevich} of the above Langevin
equations is possible~\cite{footnote:phasereduction}.
Specifically, we describe the dynamics of each oscillator using only a
constantly-increasing phase variable $\phi(t) = \phi({\bm X}(t)) \in
[-\pi, \pi]$, defined along its limit cycle and also on its phase
space except at phaseless sets.
Applying the standard phase reduction method to
Eq.~(\ref{lan0})~\cite{Kuramoto}, we obtain (by virtue of the
Stratonovich interpretation) the following approximate Langevin
equations for the phase variables ${\bm \phi} = ( \phi^{(1)}, \cdots,
\phi^{(N)} )$:
\begin{equation}
  \dot{\phi}^{(\alpha)}(t) = \omega
  + \sqrt{D} {\bm Z}(\phi^{(\alpha)}) \cdot
  {\bm G}(\phi^{(\alpha)}) {\bm \xi}(t)
  + \sqrt{\epsilon} {\bm Z}(\phi^{(\alpha)}) \cdot
  {\bm H}(\phi^{(\alpha)}) {\bm \eta}^{(\alpha)}(t).
  \label{phase1}
\end{equation}
Here, $\omega$ is the natural frequency of the oscillators, ${\bm
  Z}(\phi^{(\alpha)}) = \nabla_{\bm X} \phi^{(\alpha)} |_{{\bm X} =
  {\bm X}_0(\phi^{(\alpha)})} \in {\bm R}^M$ is the phase sensitivity
function of the individual oscillator that quantifies the phase
response of each oscillator to weak perturbations~\cite{Kuramoto},
${\bm G}(\phi^{(\alpha)}) = {\bm G}({\bm X}_0(\phi^{(\alpha)}))$, and
${\bm H}(\phi^{(\alpha)}) = {\bm H}({\bm X}_0(\phi^{(\alpha)}))$.
We normalize ${\bm Z}(\phi)$ such that ${\bm Z}(\phi) \cdot {\bm
  F}({\bm X}_0(\phi)) \equiv \omega$ holds constantly.
${\bm Z}(\phi)$, ${\bm G}(\phi)$, and ${\bm H}(\phi)$ are smooth
periodic functions of $\phi$.

The Stratonovich Langevin equations~(\ref{phase1}) are converted to
equivalent Ito stochastic differential equations~\cite{SDE} of the
form $ d\phi^{(\alpha)}(t) = A^{(\alpha)}({\bm \phi}) dt + d
\zeta^{(\alpha)}({\bm \phi}, t) $,
where $\{ \zeta^{(\alpha)}({\bm \phi}, t) \}$ are correlated Wiener
processes. Their increments are expressed as
\begin{equation}
  d \zeta^{(\alpha)}({\bm \phi}, t) = \sqrt{D} \sum_{k=1}^{M} \left(
    \sum_{i=1}^{M} Z_{i}(\phi^{(\alpha)}) G_{i k}(\phi^{(\alpha)})
  \right) d V_{k}(t) + \sqrt{\epsilon} \sum_{k=1}^{M} \left(
    \sum_{i=1}^{M} Z_{i}(\phi^{(\alpha)}) H_{i k}(\phi^{(\alpha)})
  \right) d W^{(\alpha)}_{k}(t),
\end{equation}
where $\{ V_{i}(t) \}$ and $\{ W^{(\alpha)}_{i}(t) \}$ are independent
Wiener processes.
The statistics of $\{ d\zeta^{(\alpha)}({\bm \phi}, t) \}$ are
specified by $\langle d\zeta^{(\alpha)}({\bm \phi}, t) \rangle = 0$
and $\langle d\zeta^{(\alpha)}({\bm \phi}, t) d\zeta^{(\beta)}({\bm
  \phi}, t) \rangle = C^{(\alpha, \beta)}({\bm \phi}) dt$,
where $C^{(\alpha, \beta)}({\bm \phi})$ is a ${\bm R}^{N \times N}$
correlation matrix defined as
\begin{eqnarray}
  C^{(\alpha, \beta)}({\bm \phi})
  &=&
  D \sum_{k=1}^{M}
  \left( \sum_{i=1}^{M} Z_{i}(\phi^{(\alpha)}) G_{i k}(\phi^{(\alpha)}) \right)
  \left( \sum_{j=1}^{M} Z_{j}(\phi^{(\beta)}) G_{j k}(\phi^{(\beta)}) \right)
  \cr \cr
  &+&
  \epsilon \sum_{k=1}^{M}
  \left( \sum_{i=1}^{M} Z_{i}(\phi^{(\alpha)}) H_{i k}(\phi^{(\alpha)}) \right)
  \left( \sum_{j=1}^{M} Z_{j}(\phi^{(\beta)}) H_{j k}(\phi^{(\beta)}) \right)
  \delta_{\alpha, \beta}.
\end{eqnarray}
Note that $C^{(\alpha, \beta)}({\bm \phi})$ is periodic in
$\phi^{(\alpha)}$ for all $\alpha$, and its $(\alpha,
\beta)$-component depends only on $\phi^{(\alpha)}$ and
$\phi^{(\beta)}$.
Since $C^{(\alpha, \beta)}({\bm \phi})$ is a symmetric positive
semi-definite matrix, we can also express $d \zeta^{(\alpha)}({\bm
  \phi}, t)$ using $N$ independent Wiener processes $\{ W^{(\beta)}(t)
\}$ as $ d \zeta^{(\alpha)}({\bm \phi}, t) = \sum_{\beta=1}^{N}
B^{(\alpha, \beta)}({\bm \phi}) dW^{(\beta)}(t)$, where $B^{(\alpha,
  \beta)}({\bm \phi})$ is a real symmetric matrix satisfying
$\sum_{\nu=1}^{N} B^{(\alpha, \nu)}({\bm \phi}) B^{(\beta, \nu)}({\bm
  \phi}) = C^{(\alpha, \beta)}({\bm \phi})$.
The transformed drift coefficients $A^{(\alpha)}({\bm \phi})$ can be
calculated as
\begin{equation}
  A^{(\alpha)}({\bm \phi}) = \omega + \frac{1}{4} \frac{\partial}{\partial \phi^{(\alpha)}}
  C^{(\alpha,\alpha)}({\bm \phi}),
\end{equation}
where we
utilized the fact that the right-hand side of Eq.~(\ref{phase1})
depends only on $\phi^{(\alpha)}$ in calculating the Wong-Zakai
correction term~\cite{SDE}.
The original $N$ vector Stratonovich Langevin equations~(\ref{lan0})
with $N+1$ independent vector noises ${\bm \xi}(t)$ and $\{ {\bm
  \eta}^{(\alpha)}(t) \}$ are now reduced to $N$ scalar Ito stochastic
differential equations with $N$ correlated scalar noises $\{ d
\zeta^{(\alpha)}({\bm \phi}, t) \}$.
The corresponding Fokker-Planck equation (FPE) describing the
evolution of the probability density function (PDF) $P({\bm \phi}, t)$
of the phase variables is given by~\cite{SDE}
\begin{equation}
  \frac{\partial}{\partial t} P({\bm \phi}, t)
  = - \sum_{\alpha=1}^{N} \frac{\partial}{\partial \phi^{(\alpha)}}
  \left( A^{(\alpha)}(\bm \phi) P \right) +
  \frac{1}{2} \sum_{\alpha=1}^{N} \sum_{\beta=1}^{N} \frac{\partial^2}
  {\partial \phi^{(\alpha)} \partial \phi^{(\beta)}} \left( C^{(\alpha,
      \beta)}(\bm \phi) P \right).
\end{equation}

We now invoke the averaging approximation~\cite{Kuramoto} to this
FPE.
We introduce new slow phase variables ${\bm \psi} = ( \psi^{(1)},
\cdots, \psi^{(N)} )$ as $\phi^{(\alpha)} = \omega t +
\psi^{(\alpha)}$ $(\alpha = 1, \cdots, N) $, and their PDF
\begin{equation}
  Q({\bm
    \psi}, t) = Q(\{ \psi^{(\alpha)} \}, t) = P(\{ \phi^{(\alpha)} =
  \omega t + \psi^{(\alpha)} \}, t).
\end{equation}
With sufficiently weak external noises, $Q$ varies slowly compared
with the oscillator natural period, $T = 2\pi / \omega$.
We can thus average the drift coefficients $A^{(\alpha)}({\bm \phi})$
and the diffusion coefficients $C^{(\alpha, \beta)}({\bm \phi})$ of
the FPE over the period $T$ keeping $Q$ constant.
The resulting averaged FPE for $Q$ is given by
\begin{equation}
  \frac{\partial}{\partial t} Q({\bm \psi}, t)
  =
  \frac{1}{2} \sum_{\alpha=1}^{N} \sum_{\beta=1}^{N}
  \frac{\partial^2}{\partial \psi^{(\alpha)} \partial \psi^{(\beta)}}
  \left( D^{(\alpha, \beta)}({\bm \psi}) Q \right).
  \label{fpe2}
\end{equation}
The drift coefficient $A^{(\alpha)}({\bm \phi})$ simply yields
$\omega$ after averaging due to the periodicity of $C^{\alpha,
  \beta}({\bm \phi})$ in $\phi^{(\alpha)}$, which vanishes in the new
variables.
The averaged diffusion coefficients $D^{(\alpha, \beta)}({\bm \psi})$
are given by
\begin{eqnarray}
  D^{(\alpha, \beta)}({\bm \psi})
  &=& \frac{1}{T} \int_{t}^{t+T}
  C^{(\alpha, \beta)}( \{ \phi^{(\alpha)} = \omega t' + \psi^{(\alpha)} \} ) dt'
  =
  D g(\psi^{(\alpha)} - \psi^{(\beta)}) + \epsilon h(0) \delta_{\alpha, \beta},
\end{eqnarray}
where we utilized the fact that $C^{(\alpha, \beta)}({\bm \phi})$
depends only on $\phi^{(\alpha)}$ and $\phi^{(\beta)}$, and introduced
the correlation function $g(\theta)$ of $Z_{i}(\phi)$ and $G_{i
  k}(\phi)$ as
\begin{eqnarray}
  g(\theta) &=& \frac{1}{2 \pi} \int_{-\pi}^{\pi} \sum_{i, j, k=1}^{M} 
  Z_{i}(\phi') G_{i k}(\phi')
  Z_{j}(\phi' + \theta) G_{j k}(\phi' + \theta) d\phi',
  \label{correlation}
\end{eqnarray}
and similarly the correlation function $h(\theta)$ of $Z_{i}(\phi)$
and $H_{i k}(\phi)$ as
\begin{eqnarray}
  h(\theta) &=& \frac{1}{2 \pi} \int_{-\pi}^{\pi} \sum_{i, j, k=1}^{M} 
  Z_{i}(\phi') H_{i k}(\phi')
  Z_{j}(\phi' + \theta) H_{j k}(\phi' + \theta) d\phi'.
  \label{correlation}
\end{eqnarray}

Clearly, $g(0) > 0$ and $h(0) > 0$ (we exclude the non-physical case
${\bm Z}(\phi) \equiv \mbox{const.}$).
Using the periodicity of ${\bm Z}(\phi)$ and ${\bm G}(\phi)$ in
$\phi$, it can also be proven that $g(\theta) = g(-\theta)$ and $g(0)
\geq g(\theta)$.
Since ${\bm Z}(\phi)$ and ${\bm G}(\phi)$ are smooth functions of
$\phi$, $g(\theta)$ has a quadratic peak at $\theta = 0$.
It can also have other quadratic peaks at $\theta \neq 0$, e.g.
$\theta = \pm 2 \pi / 3$, when ${\bm Z}(\phi)$ contains non-negligible
high-order harmonics or when the common noise is introduced
multiplicatively.

To analyze the phase relationship between the oscillators, we focus on
the PDF of the phase difference.
Without loss of generality, we first introduce the two-body PDF of
$\psi^{(1)}$ and $\psi^{(2)}$ as $ R(\psi^{(1)}, \psi^{(2)}, t) = \int
d\psi^{(3)} \cdots d\psi^{(N)} Q({\bm \psi}, t) $.
The evolution equation for $R(\psi^{(1)}, \psi^{(2)}, t)$ can be
derived from Eq.(\ref{fpe2}) by integrating over all other phase
variables as
\begin{eqnarray}
  \frac{\partial}{\partial t} R(\psi^{(1)}, \psi^{(2)}, t)
  &=&
  \frac{1}{2} \left( D g(0) + \epsilon h(0) \right)
  \left\{
    \left(\frac{\partial}{\partial \psi^{(1)}}\right)^2
    +
    \left(\frac{\partial}{\partial \psi^{(2)}}\right)^2
  \right\}
  R
  +
  \frac{\partial^2}{\partial \psi^{(1)} \partial \psi^{(2)}}
  \left(
    D g(\psi^{(1)} - \psi^{(2)}) R
  \right).
  \;\;\; \;\;\;
  \label{fpe3}
\end{eqnarray}
Furthermore, by transforming the two phase variables to the mean phase
and the phase difference, $\psi = ( \psi^{(1)} + \psi^{(2)} ) / 2$,
$\theta = \psi^{(1)} - \psi^{(2)}$, the above equation can be further
decoupled as
\begin{eqnarray}
  \frac{\partial}{\partial t} S(\psi, t)
  &=&
  \frac{1}{4} \left\{ D [ g(0) + g(\theta) ] + \epsilon h(0) \right\} 
  \frac{\partial^2}{\partial \psi^{2}} S(\psi, t),
  \cr\cr
  \frac{\partial}{\partial t} U(\theta, t)
  &=&
  \frac{\partial^2}{\partial \theta^{2}}
  \left\{ D [ g(0) - g(\theta) ] + \epsilon h(0) \right\} 
  U(\theta, t),
  \label{fpe4}
\end{eqnarray}
where $S(\psi, t) U(\theta, t) = R( \psi^{(1)}=\psi+\theta/2,
\psi^{(2)}=\psi-\theta/2, t )$.
It is clear that Eq.~(\ref{fpe4}) has a unique final stationary state,
where the PDF of the mean phase $\psi$ is uniform over the limit
cycle, $S_{0}(\psi) \equiv 1 / 2 \pi$, and the PDF of the phase
difference $\theta$ is given by
\begin{equation}
  U_{0}(\theta)
  = \frac{u_0}{ D[ g(0) - g(\theta) ] + \epsilon h(0) },
  \label{stapdf}
\end{equation}
where $u_0$ is a normalization constant.

We now examine the consequences of the above results.
Our argument holds generally for arbitrary $g(\theta)$ that satisfies
our assumptions, namely, for a general class of limit-cycle
oscillators.
When only the independent noises are given, $D=0$ and $\epsilon>0$,
$U_{0}(\theta)$ is simply uniform, so that the oscillators are
completely desynchronized.
When only the common noise is given, $D>0$ and $\epsilon=0$,
$U_{0}(\theta)$ diverges at $\theta = 0$ while remaining positive
because $g(0) \geq g(\theta)$, so that the phase difference between
any pair of oscillators accumulates at zero, resulting in
noise-induced complete synchronization.
As $\epsilon$ is increased from zero, $U_{0}(\theta)$ becomes broader,
but its peak at $\theta = 0$ remains as long as $D > 0$, i.e. the
oscillators still concentrate coherently around $\theta = 0$.
As we mentioned previously, $g(\theta)$ may have multiple peaks in
addition to $\theta = 0$. Then, the above discussion also holds for
such values of $\theta$.
Multiple peaks of $g(\theta)$ lead to the clustering behavior of the
oscillators, a well-known phenomenon in coupled
oscillators~\cite{Clustering}, but in the present case, it is caused
by the combined effect of the phase sensitivity and the common noise
alone.
More generally, $U_{0}(\theta)$ can exhibit a wide variety of
non-uniform ``coherent'' distributions depending on the functional
form of $g(\theta)$.

We can also examine the statistical stability of the synchronized
state $\theta = 0$ and the dynamics of $\theta$ around it.
From Eq.(\ref{fpe4}), we obtain the corresponding Ito stochastic
differential equation for $\theta$ as $d\theta(t) = \sqrt{2} \left\{ D
  [ g(0) - g(\theta) ] + \epsilon h(0) \right\}^{\frac{1}{2} } dw(t)$,
where $w(t)$ is a Wiener process.
Focusing on the region around $\theta = 0$, we approximate $g(\theta)$
around its $\theta = 0$ peak as $g(\theta) \simeq g(0) - (1/2)
|g''(0)| \theta^2$, utilizing the facts that $g'(0) = 0$ and $g''(0) <
0$, where $'$ denotes $d / d\theta$.
We then obtain
\begin{equation}
  d\theta(t) \simeq \sqrt{ D |g''(0)| } \theta(t)
  dw_1(t) + \sqrt{ 2 \epsilon h(0) } dw_2(t),  
\end{equation}
where the noise is
decomposed into multiplicative and additive parts using two
independent Wiener processes $w_{1, 2}(t)$.
This is simply a linear random multiplicative process with an additive
noise~\cite{OnOff}.
Let us ignore the additive noise $dw_{2}(t)$ for the moment.  Using
the Ito formula~\cite{SDE}, the equation for the logarithm of the
absolute phase difference is obtained as $d\ln|\theta(t)| = -
\frac{1}{2} D |g''(0)| dt + \sqrt{ D |g''(0)| } dw_1(t)$,
so that the average Lyapunov exponent of the completely synchronized
state $\theta = 0$ is given by $ \lambda = - \frac{1}{2} D |g''(0)| <
0$, which is always negative, i.e. $\theta = 0$ is always
statistically stable.
When the common noise is additive, ${\bm G}(\phi)$ is a constant
matrix, and we recover the previous results~\cite{Teramae,Goldobin}.
When weak independent noises exist, $|\theta|$ mostly remains small
but occasionally exhibits large bursts, a typical behavior known as
noisy on-off intermittency~\cite{OnOff}.
We then expect a power-law PDF of the inter-burst intervals of
$\theta(t)$ with an exponent $-1.5$, and also a power-law PDF of the
phase differences around $\theta = 0$, whose exponent is always $-2$
in the present case~\cite{OnOff} (results not shown; see
Ref.~\cite{Teramae}).
When $g(\theta)$ has multiple peaks, we can estimate the stability and
fluctuations around the other peaks in a similar fashion, and we
expect intermittent transitions between the clustered
states~\cite{footnote:intermittency}.

\begin{figure}[!htbp]
  \begin{center}
    \includegraphics[width=1.0\hsize,clip]{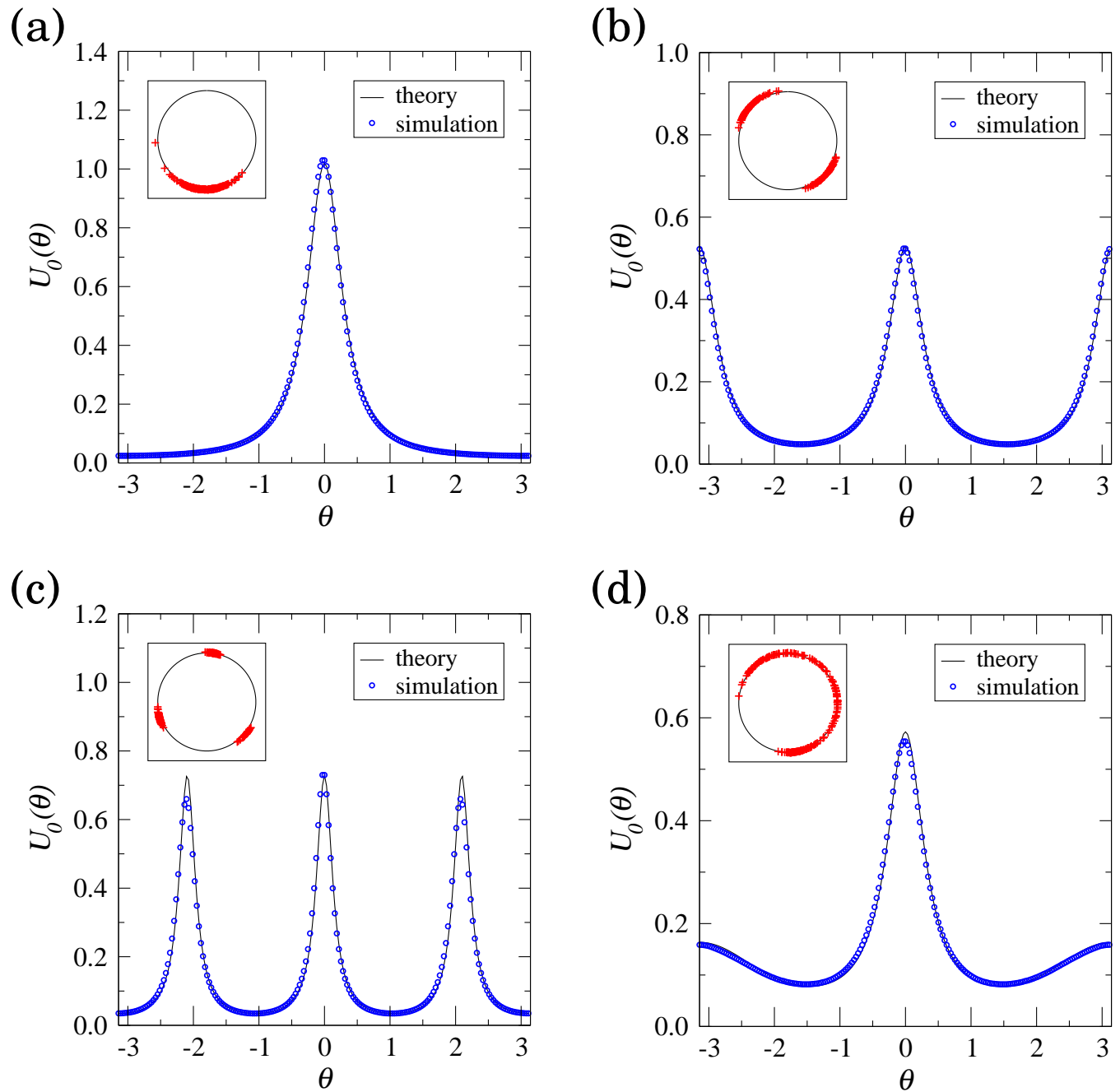}
    \caption{Stuart-Landau oscillators. (a) Synchronized, (b)
      2-cluster, (c) 3-cluster, and (d) intermixed states.  The
      insets display instantaneous distributions of the oscillators
      on the limit cycle.}
    \label{fig1}
  \end{center}
\end{figure}

We now demonstrate the noise-induced synchronization and clustering
numerically.
As the first example, we consider uncoupled Stuart-Landau (SL)
oscillators, ${\bm X} = \left(x, y\right)$, ${\bm F}({\bm X}) = (\; x
- c_0 y - (x^2 + y^2) (x - c_2 y),\; y + c_0 x - (x^2 + y^2) (y + c_2
x) \;)$,
subject to independent additive noises,
${\bm H}({\bm X}) = \mbox{diag}(1, 1)$,
and to the following four types of additive or multiplicative common
noises,
${\bm G}_1({\bm X}) = \mbox{diag}(1, 1)$, ${\bm G}_2({\bm X}) =
\mbox{diag}(x, y)$, ${\bm G}_3({\bm X}) = \mbox{diag}(1 + 4 xy, 0)$,
and ${\bm G}_4({\bm X}) = \mbox{diag}(x, xy)$.
The SL oscillator is the simplest limit-cycle oscillator derived as a
normal form of the supercritical Hopf bifurcation~\cite{Kuramoto}.
We fix the parameters at $c_0 = 2$ and $c_2 = -1$, with which the
natural frequency becomes $\omega = c_0 - c_2 = 3$.  The phase
sensitivity function is analytically given as ${\bm Z}(\phi) =
\sqrt{2} (\; \sin \left( \phi + 3 \pi / 4 \right), \sin \left( \phi +
  \pi / 4 \right) \;)$~\cite{Kuramoto}.
From Eq.(\ref{correlation}), we obtain the corresponding correlation
functions as $g_1(\theta) = 2 \cos \theta$, $g_2(\theta) = \cos^2
\theta$, $g_3(\theta) = \cos 3\theta$, $g_4(\theta) = ( \cos \theta +
8 \cos^2 \theta + \cos 3\theta ) / 16$, and $h(0) = 2$, from which we
can calculate $U_0(\theta)$.
We thus expect noisy synchronization (1-cluster), 2-cluster,
3-cluster, and intermixed coherent distributions of $\theta$ to be
observed.
Figure \ref{fig1} compares the results of direct numerical simulations
using $N = 200$ oscillators with the analytical results, where the
noise intensities are fixed at $D = 0.002$ and $\epsilon = 0.0001$.
To realize the Stratonovich situation, the numerical simulations are
performed using colored Gaussian white noises generated by the
Ornstein-Uhlenbeck process $\tau \dot{z}(t) = - z(t) + \xi(t) $ with a
small correlation time $\tau = 0.05$, where $\xi(t)$ is a Gaussian
white noise of unit intensity~\cite{SDE}.
As expected, various synchronized or clustered states are realized,
and their PDFs are fitted nicely by the theoretical curves.

\begin{figure}[!htbp]
  \begin{center}
    \includegraphics[width=1.0\hsize,clip]{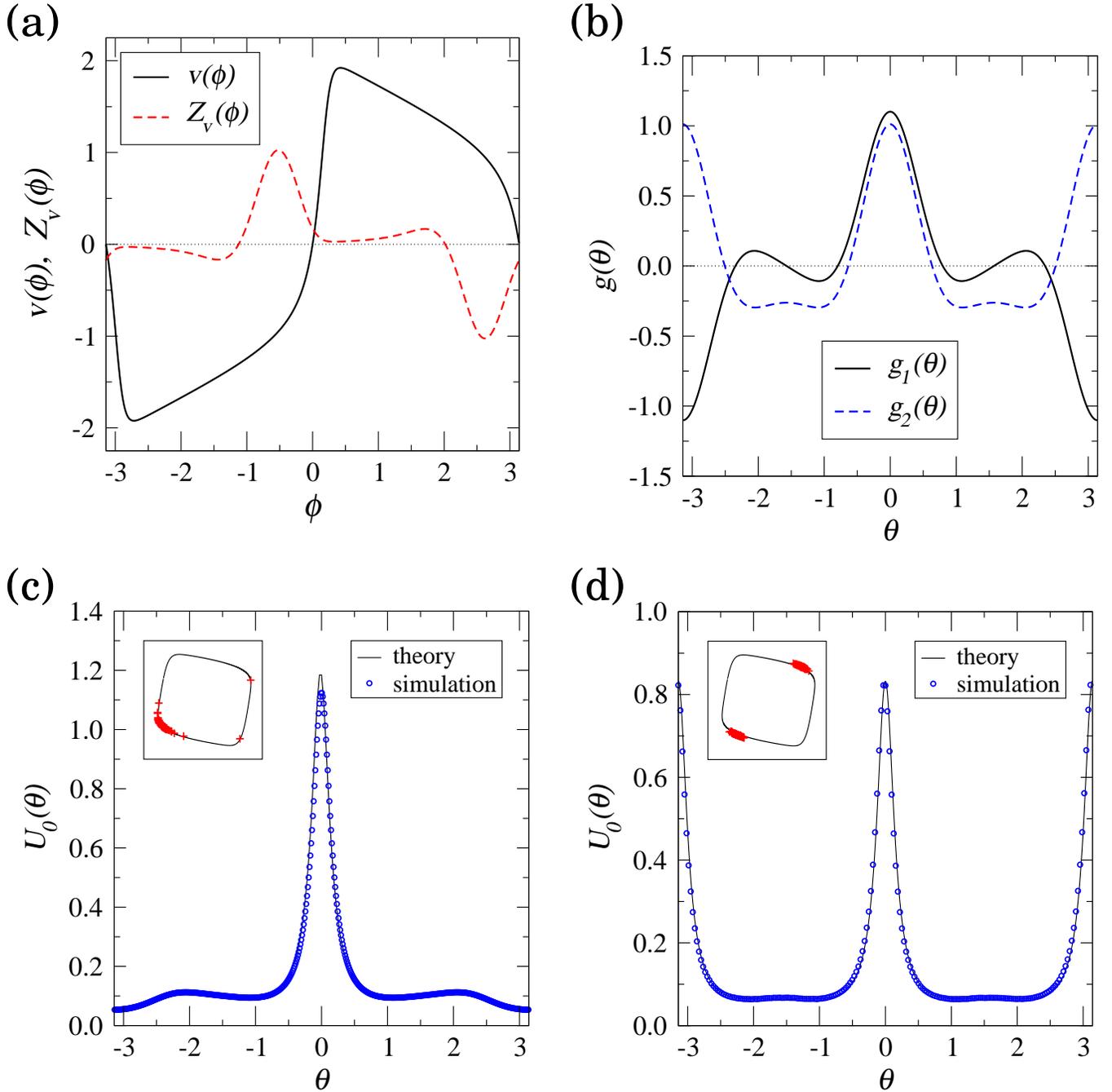}
    \caption{FitzHugh-Nagumo oscillators. (a) Variable $v(\phi)$ and
      phase sensitivity function $Z_v(\phi)$, (b) correlation
      functions $g_{1,2}(\theta)$ calculated from ${\bm G}_{1,2}({\bm
        X})$ and $Z_v(\phi)$, (c) synchronized state, and (d)
      2-cluster state.  The insets display snapshots of the
      oscillators.}
    \label{fig2}
  \end{center}
\end{figure}

As the second example, we consider uncoupled FitzHugh-Nagumo (FN)
oscillators~\cite{Izhikevich},
${\bm X} = ( u, v )$, ${\bm F}({\bm X}) = (\; \varepsilon ( v + c - d
u ), v - v^3 / 3 - u + I \;)$, subject to either an additive or
multiplicative common noise,
${\bm G}_1({\bm X}) = \mbox{diag}(0, 1)$ or ${\bm G}_2({\bm X}) =
\mbox{diag}(0, v)$,
and also to an additive independent noise, ${\bm H}({\bm X}) =
\mbox{diag}(0, 1)$.
The noises are applied only to the variable $v$ corresponding to the
membrane potential.
Fixing the parameter values at $\varepsilon = 0.08$, $c = 0.7$, $d =
0.8$, and $I = 0.875$, the limit cycle becomes symmetric
with a natural frequency of $\omega \simeq 0.1725$.
The phase sensitivity function $Z_v(\phi)$ can be calculated
numerically using the method devised in~\cite{Ermentrout,Izhikevich}.
Figure \ref{fig2} compares the results of direct numerical simulations
with the analytical results at $D = 0.005$ and $\epsilon = 0.0005$
using $N = 200$ oscillators.
Either synchronized or $2$-cluster states are realized for the
additive or multiplicative common noises, and their PDFs are well
fitted by the theoretical curves calculated using the numerical
$Z_v(\phi)$.

Summarizing, we developed a global formulation of synchronization
and clustering phenomena in ensembles of uncoupled limit-cycle
oscillators induced by a common noise.
The common noise acts as a state-dependent noise on the phase
difference, which yields the $\theta$-dependent effective diffusion
constant for $U(\theta, t)$ in Eq.~(\ref{fpe4}), and results in the
non-uniform coherent stationary distribution $U_{0}(\theta)$ in
Eq.~(\ref{stapdf}).
In our formulation, the combination of the common and independent
noises is a special case of more general correlated noises, and the
synchronized or clustered state is a special case of non-uniform
coherent distributions.
Thus, we can generalize the notion of {\it common-noise-induced
  synchronization} to {\it correlated-noise-induced coherence}.
This insight will be helpful in understanding various noise-induced
synchronization phenomena.

We thank Y. Kuramoto for useful comments, and the Grant-in-Aid for the
21st Century COE ``Center for Diversity and Universality in Physics''
from the Ministry of Education, Culture, Sports, Science and
Technology of Japan for financial support.

\begin{thebibliography}{99}
\bibitem{footnote:dispersion} When only concerned with overall
  distribution of the phase differences, effect of small dispersion in
  oscillator characteristics is qualitatively similar to introduction
  of independent noises in Eq.~(\ref{lan0}).  Here, we do not treat
  such dispersion for simplicity.
\bibitem{footnote:phasereduction} For Gaussian processes,
  finite-distance jump probability in interval $\Delta t$ goes to $0$
  faster than $\Delta t$ as $\Delta t \to 0$ (Lindeberg
  condition~\cite{SDE}), so as long as noise intensities $D$ and
  $\epsilon$ are sufficiently small, the trajectory of each oscillator
  is almost always near unperturbed limit cycle, so that phase
  approximation Eq.~(\ref{phase1}) is valid.
\bibitem{footnote:intermittency} This intermittent transition is also
  induced by the noise, which is different from the dynamical
  intermittency caused by the intricate phase-space structure analyzed
  in~\cite{Clustering}, and thus exhibits different statistics.
\end{thebibliography}
\end{document}